\documentclass[journal=jacsat,manuscript=article]{achemso}

\usepackage[T1]{fontenc}
\usepackage[nomath]{lmodern}
\usepackage[utf8]{inputenc}

\usepackage[dvipsnames]{xcolor}
\definecolor{lightblue}{RGB}{51,102,187}
\usepackage[linkcolor = blue, citecolor= lightblue]{hyperref}

\usepackage[super]{natbib}
\hypersetup{colorlinks=false}
\usepackage[english]{babel}
\usepackage{gensymb}
\usepackage{amssymb}
\usepackage{float}
\usepackage{xspace} 
\usepackage[labelfont=bf]{caption}
\usepackage[figurename=Fig.]{caption}
\usepackage{lineno}
\usepackage[version=3]{mhchem} 

\author{Sambit Ghosh}
\affiliation{Univ. Grenoble Alpes, CEA, CNRS, Grenoble INP, IRIG-Spintec, 38054 Grenoble, France}
\alsoaffiliation{Institute of Applied Physics, Graduate School of Pure and Applied Sciences, University of Tsukuba, Tsukuba, Ibaraki 305–8573, Japan}
\author{Taro Komori}
\affiliation{Institute of Applied Physics, Graduate School of Pure and Applied Sciences, University of Tsukuba, Tsukuba, Ibaraki 305–8573, Japan}
\author{Ali Hallal}
\affiliation{Univ. Grenoble Alpes, CEA, CNRS, Grenoble INP, IRIG-Spintec, 38054 Grenoble, France}
\author{Jose Pe\~{n}a  Garcia}
\affiliation{Univ. Grenoble Alpes, CNRS, Institut N\'eel, 38042 Grenoble, France}
\author{Toshiki Gushi}
\affiliation{Institute of Applied Physics, Graduate School of Pure and Applied Sciences, University of Tsukuba, Tsukuba, Ibaraki 305–8573, Japan}
\alsoaffiliation{Univ. Grenoble Alpes, CEA, CNRS, Grenoble INP, IRIG-Spintec, 38054 Grenoble, France}
\author{Taku Hirose}
\affiliation{Institute of Applied Physics, Graduate School of Pure and Applied Sciences, University of Tsukuba, Tsukuba, Ibaraki 305–8573, Japan}
\author{Haruka Mitarai}
\affiliation{Institute of Applied Physics, Graduate School of Pure and Applied Sciences, University of Tsukuba, Tsukuba, Ibaraki 305–8573, Japan}
\author{Hanako Okuno}
\affiliation{Univ. Grenoble Alpes, CEA, IRIG-MEM, 38000 Grenoble, France}
\author{Jan Vogel}
\affiliation{Univ. Grenoble Alpes, CNRS, Institut N\'eel, 38042 Grenoble, France}
\author{Mairbek Chshiev}
\affiliation{Univ. Grenoble Alpes, CEA, CNRS, Grenoble
INP, IRIG-Spintec, 38054 Grenoble, France}
\alsoaffiliation{Institut Universitaire de France, 75231, Paris, France}
\author{Jean-Philippe Attan\'e}
\affiliation{Univ. Grenoble Alpes, CEA, CNRS, Grenoble INP, IRIG-Spintec, 38054 Grenoble, France}
\author{Laurent Vila}
\affiliation{Univ. Grenoble Alpes, CEA, CNRS, Grenoble INP, IRIG-Spintec, 38054 Grenoble, France}
\email{laurent.vila@cea.fr}
\author{Takashi Suemasu}
\affiliation{Institute of Applied Physics, Graduate School of Pure and Applied Sciences, University of Tsukuba, Tsukuba, Ibaraki 305–8573, Japan}
\author{Stefania Pizzini}
\affiliation{Univ. Grenoble Alpes, CNRS, Institut N\'eel, 38042 Grenoble, France}
\email{stefania.pizzini@neel.cnrs.fr}

\title
  {Current-driven domain wall dynamics in ferrimagnetic Ni-doped Mn$_{4}$N films : very large domain wall velocities and reversal of motion direction across the magnetic compensation point}

\abbreviations{DW, STT,SOT,SHE,AHE, MRAM}
\DeclareUnicodeCharacter{2212}{x}

\begin{document}

\begin{abstract}
 Spin-transfer torque (STT) and spin-orbit torque (SOT)  are spintronic phenomena allowing magnetization manipulation using electrical currents. Beyond their fundamental interest, they allow developing new classes of magnetic memories and logic devices, in particular based on domain wall (DW) motion.
 In this work, we report the study of STT driven DW motion in ferrimagnetic manganese nickel nitride (Mn$_{4-x}$Ni$_x$N) films, in which  magnetization and angular momentum compensation can be obtained by the fine adjustment of the Ni content. Large domain wall velocities, approaching 3000 m/s,  are measured  for Ni compositions close to the angular momentum compensation point. The  reversal of the DW motion direction,  observed when the  compensation composition is crossed, is related to the change of direction of the angular momentum with respect to that of the spin polarization.  This is confirmed by the results of \emph{ab initio}  band structure calculations.
\end{abstract}


\section{Introduction}
Domain walls (DWs) separate magnetic domains present in ferromagnetic materials. Current-driven DW motion was predicted theoretically by Berger in 1978 \cite{berger_lowfield_1978} and then  extensively studied for its potential applications in domain wall racetrack memories \cite{parkin_memory_2015}, DW MRAM\cite{brataas_current-induced_2012}, spin torque majority gate \cite{nikonov_proposal_2011,vaysset_toward_2016} and other domain-wall-based logic devices \cite{allwood_submicrometer_2002,currivan_low_2012,currivan-incorvia_logic_2016,luo_current-driven_2020}. The two mechanisms leading to DW motion driven by spin polarized currents are the spin-transfer torque  (STT) and the spin orbit torque (SOT) associated to the spin Hall effect (SHE) \cite{slonczewski_current-driven_1996}. While in the case of STT the charge current is spin polarized within the ferromagnetic layer by \emph{sd} exchange interaction \cite{freitas_observation_1985,grollier_switching_2003,yamaguchi_real-space_2004}, in the case of SHE-SOT,  the spin polarized current is generated by SHE in a neighbouring film (Pt, W, Ta ..), and then injected into the ferromagnetic layer \cite{khvalkovskiy_high_2009,haazen_domain_2013,Emori2013,ryu_chiral_2013}, resulting in both cases into a torque applied on the DWs. 

Studies on systems in which domain walls are driven by STT are nowadays rare, mainly because there are practically no reports of efficient STT in thin films with perpendicular magnetization. 
In the last decade the interest of the spintronic community has focused on thin ferromagnetic films deposited on a heavy metal, where the interfacial Dzyaloshinskii-Moriya interaction (DMI) stabilizes chiral N\'eel walls that can be efficiently driven by SHE-SOT \cite{Emori2013,ryu_chiral_2013}. 

Recently, current-induced magnetization dynamics in ferrimagnets has become an active field of research. In these materials,  two magnetic sub-lattices are anti-ferromagnetically coupled, and the magnetization and angular momentum compensation of the two sublattices 
may be obtained by varying either the temperature or the composition of the material. Previous experiments on ferrimagnets \cite{kim_fast_2017,caretta_fast_2018,siddiqui_current-induced_2018,Caretta2020}, mostly carried out on thin films deposited on a heavy metal, have evidenced new physical mechanisms  taking place in the vicinity of these compensation points, leading  to large SOT-driven DW velocities \cite{kim_fast_2017,caretta_fast_2018}. 

In the present work we report the study of ferrimagnetic Mn$_4$N thin films doped with Ni, in which magnetization and angular momentum compensation can be obtained by the fine adjustment of the Ni content. 
In these films, neither bulk nor interfacial DMI are present; domain walls have then a Bloch internal structure and are driven by STT. Our Kerr microscopy measurements show that due to the relatively large spin polarization of the conduction electrons and to the reduced angular momentum  close to the compensation composition, domain walls can be driven by STT with an unprecedented efficiency. The  reversal of the DW motion direction,  observed when the   compensation composition is crossed, is related to the change of direction of the angular momentum with respect to that of the spin polarization.  This is confirmed by the results of \emph{ab initio} band structure calculations.

\section{Structural, magnetic and transport properties}

\begin{figure}[htbp]
\centering
\includegraphics[scale=0.6]{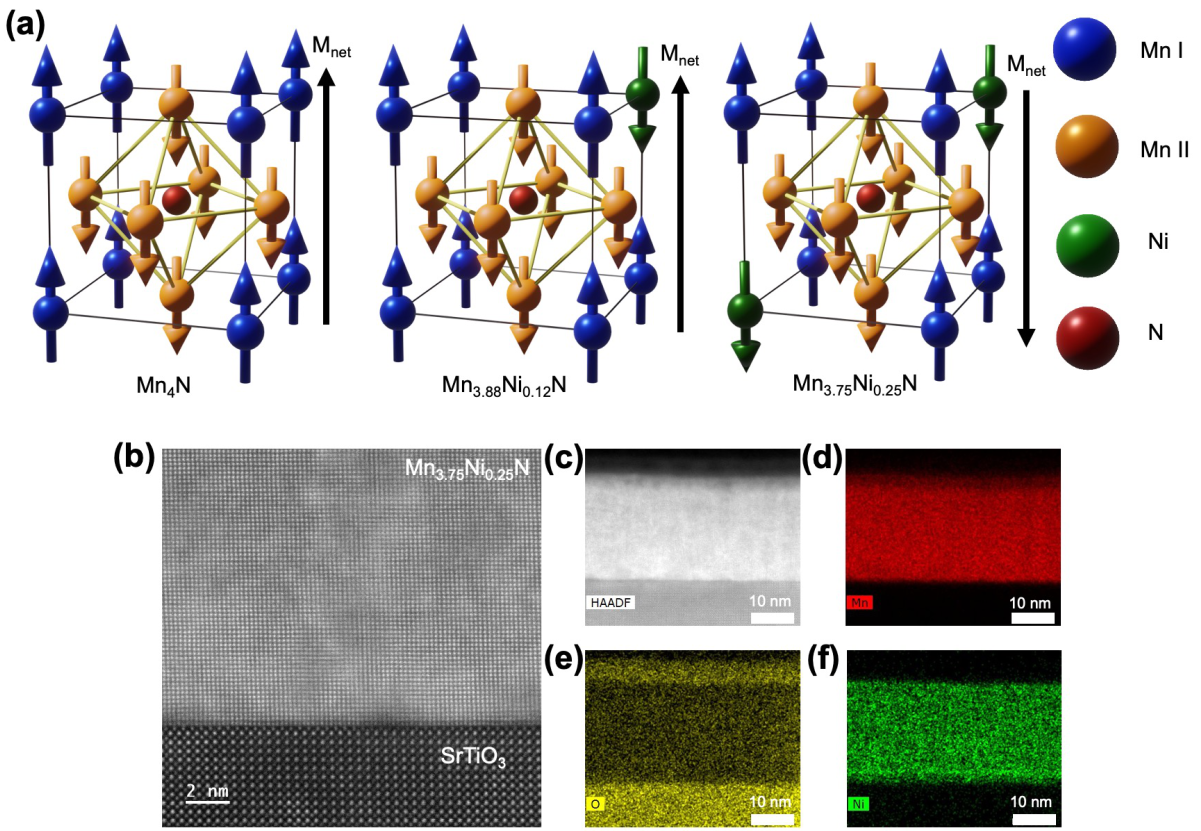}

\caption{\textbf{(a)} Schematics of the anti-perovskite crystal structure of Mn$_4$N(left), showing the substitution of the Mn(I) site atoms (in blue) with Ni atoms, to model Mn$_{1-x}$Ni$_{x}$N (middle and right). Mn(II) atoms are shown in orange, Ni atoms in green and nitrogen atoms in red. Mn(I) magnetic moments are parallel to the [001] quantization axis.  The net magnetization decreases when increasing the Ni concentration, and after crossing the magnetic compensation point the direction of the net magnetic moment is reversed. \textbf{(b)} High resolution STEM image of a 30 nm Mn$_{3.75}$Ni$_{0.25}$N thin film deposited onto a STO substrate. \textbf{(c)} HAADFSTEM image of the full thin film with the capping layer of 3 nm SiO$_2$ and associated EDX elemental map of Mn \textbf{(d)}, O \textbf{(e)} and Ni \textbf{(f)}. }
\label{fig:1}
\end{figure}

Ferrimagnetic Mn$_4$N grows with a tetragonal anti-perovskite crystal structure (Fig.~\ref{fig:1}a), with two types of Mn atoms at the corner  (site I) and at the face centred sites (site II). The two magnetic sublattices are anti-ferromagnetically coupled, with the net magnetization parallel to the Mn(I) moment.  With a high Curie temperature of 745K \cite{takei_magnetic_1960,takei_magnetic_1962,mekata_magnetic_1962,fruchart_non-collinear_1979}, a low magnetization (around 70-150 kA/m), and a relatively high uniaxial perpendicular anisotropy  (0.1 $\times$ 10$^6$ MJ/m$^3$) \cite{ching_anomalous_1994,yasutomi_perpendicular_2014,kabara_perpendicular_2015,shen_metallic_2014,meng_extrinsic_2015,ito_perpendicular_2016}, this rare-earth free material is an interesting candidate for spintronic applications.

Using either pulsed-laser deposition \cite{shen_metallic_2014}, DC reactive sputtering \cite{kabara_perpendicular_2015} or molecular beam epitaxy \cite{yasutomi_perpendicular_2014,meng_extrinsic_2015}, Mn$_4$N thin films can be grown on different substrates such as Si \cite{ching_anomalous_1994}, SiC and GaN \cite{dhar_ferrimagnetic_2005}, MgO \cite{yasutomi_perpendicular_2014,kabara_perpendicular_2015,shen_metallic_2014,meng_extrinsic_2015,ito_perpendicular_2016}, SrTiO$_3$ (STO) \cite{yasutomi_perpendicular_2014,ito_perpendicular_2016,gushi_millimeter-sized_2018} and (LaAlO$_3$)$_{0.3}$(Sr$_2$TaAlO$_6$)$_{0.7}$ (LSAT) \cite{hirose_perpendicular_2020}. Mn$_4$N films deposited on STO(001) were shown to exhibit smooth magnetic domains (at the mm scale), due to a reduced density of pinning sites  \cite{gushi_millimeter-sized_2018}, and record current-induced domain wall velocities driven by STT (900~m/s for J=1.3 TA/m$^2$) \cite{gushi_large_2019}. 

In this paper, we focus on  Ni-doped Mn$_4$N thin films epitaxially grown on STO(001) substrates \cite{komori_magnetic_2020}. 
10~nm and 30~nm thick Mn$_{4-x}$Ni$_x$N films, with nominal $x$ varying between 0.1 and 0.25, were grown by molecular beam epitaxy on STO(001) substrates and capped with a 3 nm thick SiO$_2$ layer. Scanning transmission electron microscopy (STEM), reflection high-energy electron diffraction and X-ray diffraction measurements were performed to check their crystalline quality. The high resolution STEM micrograph of a 30~nm~thick Mn$_{3.75}$Ni$_{0.25}$N film shown in Fig.~\ref{fig:1}b illustrates a highly ordered crystalline structure, with a negligible density of defects. The high-angle annular dark field scanning transmission electron microscopy (HAADF-STEM) image of the full stack and the elemental maps of Mn, O and Ni obtained by energy-dispersive X-ray spectroscopy (EDX), are shown in Fig.~\ref{fig:1}c-f. These data demonstrate that the distribution of the Ni atoms throughout the film has a good uniformity. On the other hand, while oxygen is mostly concentrated in the substrate and in the capping layer, a small region of the film, at the interface with the capping layer, appears to be oxidized. This layer has to be considered as a magnetically dead layer. 

Our previous  X-ray magnetic circular dichroism (XMCD) measurements \cite{komori_magnetic_2020}  showed that Ni occupies preferentially  the Mn(I) sites of Mn$_{4}$N.  Since the moment carried by Ni atoms is anti-parallel to that of Mn(I), increasing the Ni atomic content allows reducing the overall magnetization.  Beyond a critical Ni content, the net magnetization direction is thus expected to reverse with respect to the original one, i.e. to become anti-parallel to the Mn(I) magnetization. An analytical calculation of M$_s$ using the magnetic moments of Mn(I), Mn(II) and Ni atoms extracted from neutron diffraction measurements \cite{takei_magnetic_1960,fruchart_non-collinear_1979} predicts the magnetization compensation to occur for $x$ = 0.18, corresponding to 3.6 at\% of Ni. The presence of a magnetization compensation point around this Ni content was proved by XMCD measurements and confirmed by the sign reversal of  the anomalous Hall angle for samples with Ni content on either side of it \cite{komori_magnetic_2019}. 
The effect of the substitution of  Mn(I) atoms with  Ni atoms on the net magnetization is sketched in Fig.~\ref{fig:1}a: the net magnetization is parallel to the Mn(I) moment below the compensation point ($x$ = 0, 0.12) and becomes parallel to the Mn(II) moment above it ($x$ = 0.25). 

\begin{figure}[htbp]
    \centering
    \includegraphics[scale=0.55]{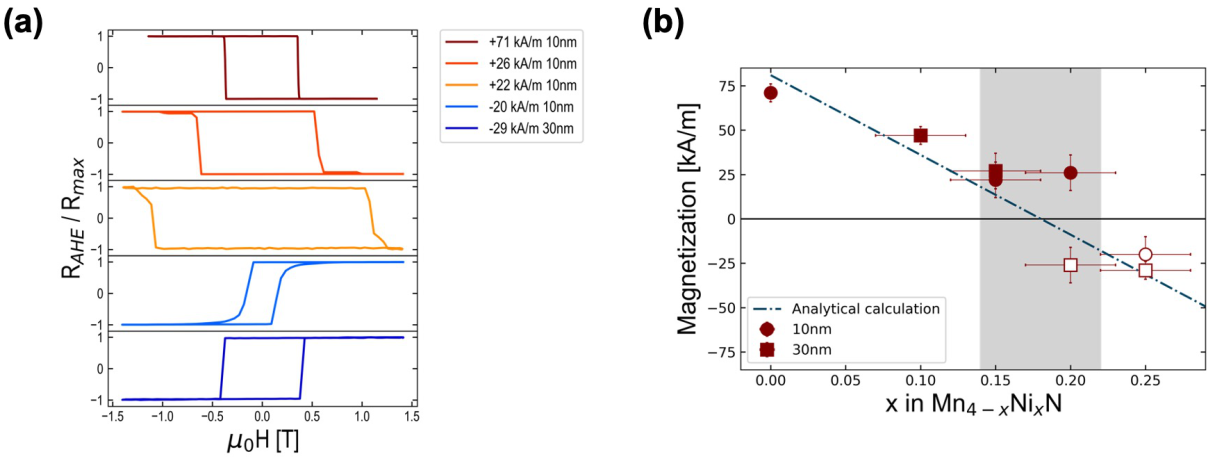}
    \caption{\textbf{(a)} Anomalous Hall effect curves measured for thin films with different Ni concentrations (x=0 (10~nm), 0.15(10~nm), 0.2(10~nm),  0.25(10~nm) and 0.25(30~nm)). The sign  of the Hall angle changes from negative to positive when crossing the magnetic compensation point between x=0.15 and x=0.25, indicating the change of direction of the net magnetization.  The corresponding net magnetization values measured by VSM-SQUID are reported in the caption. \textbf{(b)} Spontaneous magnetization versus nominal Ni concentration x, measured for the Mn$_{4-x}$Ni$_x$N thin films by VSM-SQUID. The sign of the net magnetization is obtained from the sign of the AHE angle. The shaded area indicates the spread of the observed deviation of the Ni content x at the compensation point, with respect to the nominal value. The dotted line  shows the magnetization obtained analytically.}
    \label{fig:2}
\end{figure}

The anomalous Hall effect (AHE) curves measured for Mn$_{4-x}$Ni$_x$N films patterned into Hall crosses and Van der Pauw method on blanket layers 
are shown in Fig.~\ref{fig:2}a as a function of the Ni concentration ($x$ = 0, 0.15,  0.2 and 0.25). The sharp magnetization switching observed for all the films reveals that they retain the large perpendicular magnetic anisotropy after patterning.  A switch of the AHE angle from negative to positive is observed  for  Ni content between $x$ = 0.15 and $x$ = 0.2. 
This shows that the relative orientation of spin polarization and magnetic moment changes between these two values, confirming the previously obtained compensation composition for $x$ $\approx$ 0.18. 

Fig.~\ref{fig:2}b shows the variation of the spontaneous magnetization M$_s$ as a function of the Ni content, measured by  vibrating sample magnetometry (VSM-SQUID) at room temperature. For each sample, the direction of the net magnetization is deduced from the sign of the AHE angle (as in Fig.~\ref{fig:2}a): positive/negative sign of M$_s$ in the Fig.~\ref{fig:2}b  indicates net magnetization parallel/anti-parallel to the Mn(I) magnetic moment.  In agreement with previous results \cite{komori_magnetic_2019,komori_magnetic_2020},  M$_s$ is observed to decrease with increasing Ni content and to change direction for $x$ between 0.15 and 0.2. Note that for the nominal Ni content $x$ = 0.15, a difference in the net magnetization direction is observed between the 10~nm and 30~nm thick films. This might be due to a small deviation from the targeted 3 at\% of Ni content in the two films. For this reason, in Fig.~\ref{fig:2}b the shaded area indicates the deviation from the nominal Ni content $x$ at the compensation point observed in this series of samples.

\section{Domain wall dynamics}

\begin{figure}[htbp]
    \centering
    \includegraphics[scale=0.55]{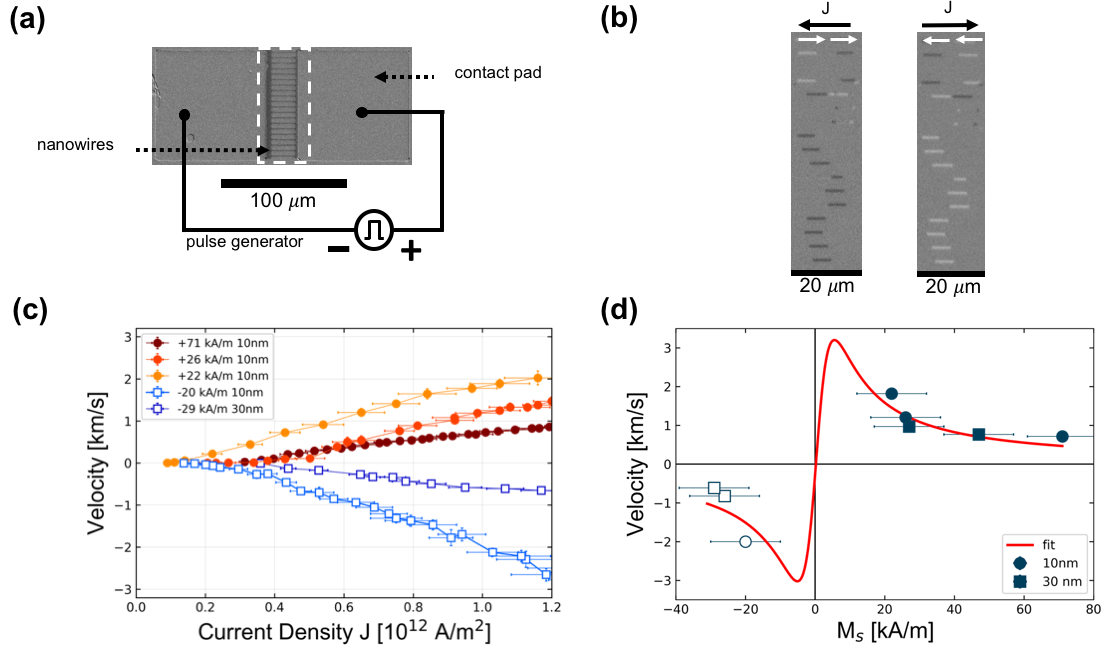}
    \caption{\textbf{(a)} Sketch of the devices fabricated  for the measurement of domain wall dynamics, showing twenty parallel  nanowires where DWs are driven by polarized current pulses, together with the contact pads from which the DWs are injected. \textbf{(b)} Differential polar MOKE images, showing the displacement of domain walls during the application of a negative (left) and positive (right) current pulses. The white arrows indicate the DW displacement. In this device with composition below the compensation point, the DWs move in the direction of the electron flow.  \textbf{(c)} Domain wall speed versus current density for Mn$_{4-x}$Ni$_x$N films  with different Ni concentration on either sides of the compensation point. $M_{S}$=71 kA/m corresponds to Mn$_{4}$N (x=0). The filled/empty symbols show the velocity below/above  the compensation point: the DW direction of motion changes sign when crossing the compensation point. \textbf{(d)} The domain wall velocity versus net magnetization $M_{S}$, measured for J=1x10$^{12}$ A/m$^2$ (black squares) is compared with the best fit obtained using the $q-\phi$ model (Eq. \ref{Eq:1}) (red line).}  
    \label{fig:3}
\end{figure}

The  domain wall dynamics in the Mn$_{4-x}$Ni$_x$N films was studied at room temperature using polar magneto-optical Kerr effect (MOKE) microscopy, with differential imaging to enhance the magnetic contrast. The samples were patterned into 1 $\mu$m-wide strips using electron beam lithography and ion milling. The image of a complete device is shown in Fig.~\ref{fig:3}a. 
Fig.~\ref{fig:3}b shows differential MOKE images illustrating, with black and white contrasts, the displacements of the domain walls in opposite directions during the application of opposite current pulses. As expected, for a given current polarity, all the DWs move in the same direction.

Fig.~\ref{fig:3}c shows the DW velocities as a function of current density $J$, measured for samples with Ni content on both sides of the magnetization compensation composition, together with that of the undoped Mn$_{4}$N sample. For compositions below the magnetization compensation point (MCP), the  DWs move in the direction of the electron flow and their mobility ($dv/dJ$) increases as the Ni concentration increases (\emph{i.e.} as the net magnetization decreases). For compositions above the MCP, the DWs  direction of motion reverses, 
and very large velocities in the direction of the the current flow, approaching 3000~m/s, are obtained for a Ni content around  $x$ =~0.25 (M$_S \approx$~-20 kA/m) and current density $J$=~1.2 $\times$ 10$^{12}$ A/m$^2$. Away from the compensation composition the DW mobility decreases and reaches values similar to those obtained for Mn$_4$N. 
The large mobilities obtained close to the magnetic  compensation composition exceed the largest SOT-driven velocities measured in ferrimagnetic GdCo/Pt thin films, close to the angular momentum compensation at 250~K \cite{caretta_fast_2018}.

In order to explain these results, let us compare them with the predictions of the collective coordinate $q-\phi$ model,  expanded to a ferrimagnetic  system consisting of two sub-lattices "1" and "2", using effective magnetic parameters \cite{Okuno2019,Haltz2021}. 


In  Mn$_{4}$N, the two magnetic sublattices "1" and "2" are composed of the same atomic species, Mn(I) and Mn(II), while in the Mn$_{4-x}$Ni$_x$N films at most 5 at\% of Mn(I) atoms are replaced by Ni atoms.  In the absence of a precise measurement of the individual gyromagnetic factors $\gamma$, and because of the small amount of Ni, we may then assume that $\gamma_{1}=\gamma_{2}$. As a consequence, the magnetic and angular compensation points coincide. Furthermore, the strong anti-ferromagnetic coupling leads us also to consider that $\alpha_{1}=\alpha_{2}$. 
In the asymptotic precessional regime i.e. well above the critical current density $J_c$ ( $\approx$ 2 $\times$ 10$^{10}$ A/m$^2$ for  Mn$_{4}$N) \cite{gushi_large_2019}, the DW velocity driven by STT reads \cite{Okuno2019,Haltz2021}:

\begin{equation}\label{Eq:1}
\mathbf{v} = \frac{L_{S} + L_\alpha \beta}{L_{S}^2+L_{\alpha}^2} L_{S} \mathbf{u}
\end{equation}

where $L_{S}\mathbf{u}=PJh/(2e)\mathbf{e}_J$, $J\mathbf{e}_J$ is the current density, $P=P_1-P_2$ is the effective spin polarization of the current, $\alpha$ is the Gilbert damping parameter, $\beta$ characterizes the non-adiabatic contribution to the STT,  $L_{S}=(M_1-M_2)/\gamma$ is the angular momentum density and $L_{\alpha}=(\alpha/\gamma)(M_1+M_2)$. 

In Fig.~\ref{fig:3}d the experimental domain wall velocities obtained from Fig.~\ref{fig:3}c for $J$= 1 $\times$ 10$^{12}$ A/m$^2$ are plotted versus $M_{S}=(M_1-M_2)=\gamma L_{S}$. The best fit of the experimental data using Eq.\ref{Eq:1}  is obtained for $P$=0.65, $\alpha$=0.013 and $\beta$=0.002. 

The reversal of DW motion direction is expected to occur for 
$L_S=-\beta L_{\alpha}$, just below the angular compensation point. In the curve in Fig. ~\ref{fig:3}c, the velocity vanishes just below the experimental magnetization compensation. This result appears to validate our assumption that $\gamma_{1}=\gamma_{2}$, and therefore that the magnetization and angular momentum compensation coincide.

These results illustrate that the unprecedented large STT-driven DW velocities measured in Mn$_{4-x}$Ni$_{x}$N can be attributed to the increase of mobility when approaching the angular momentum compensation point (ACP), together to the relatively large net spin polarization of the conduction electrons. The reversal of the domain wall velocity predicted by the analytical model is based on the assumption that the sign of the spin polarization $P$ does not change with the angular momentum density $L_S$.  The DW motion reversal is therefore  related to the  change of relative orientation of  the net spin polarization and the angular momentum, when the ACP is crossed. 

In order to confirm the validity of this assumption, we have carried out \emph{ab initio}  band structure calculations. 

\section{\emph{Ab initio} calculations}

First principles calculations were performed in the framework of density functional theory (DFT) using spin polarized relativistic Korringa-Kohn-Rostoker (SPR-KKR) and Vienna Ab Initio Simulation packages \cite{kresse_ab_1993,kresse_efficiency_1996,kresse_efficient_1996,blochl_projector_1994,perdew_generalized_1996,kresse_ultrasoft_1999,ebert_calculating_2011,ebert_fully_2016,ebert_munich_nodate}.  
The calculations for Mn$_4$N were based on the perovskite crystal structure ($Pm\overline{3}m$ space group) with lattice parameter of 3.74~\AA~ and  collinear magnetic configuration, as shown in Fig.~\ref{fig:1}a. The total magnetic moment along the [001] quantization axis was found to be 1 $\mu_B$ (Mn I: 3.3 $\mu_B$, Mn II : -0.8 $\mu_B$) in good agreement with previous calculations  \cite{meinert_exchange_2016,isogami_contributions_2020}. 
Ni doping in Mn$_{4-X}$Ni${_x}$N was taken into account using both the supercell approach and the coherent potential approximation (CPA) as implemented in SPR-KKR code \cite{ebert_calculating_2011,ebert_fully_2016,ebert_munich_nodate} (see Methods for details). 

\begin{figure}[htbp]
    \centering
    \includegraphics[scale=0.5]{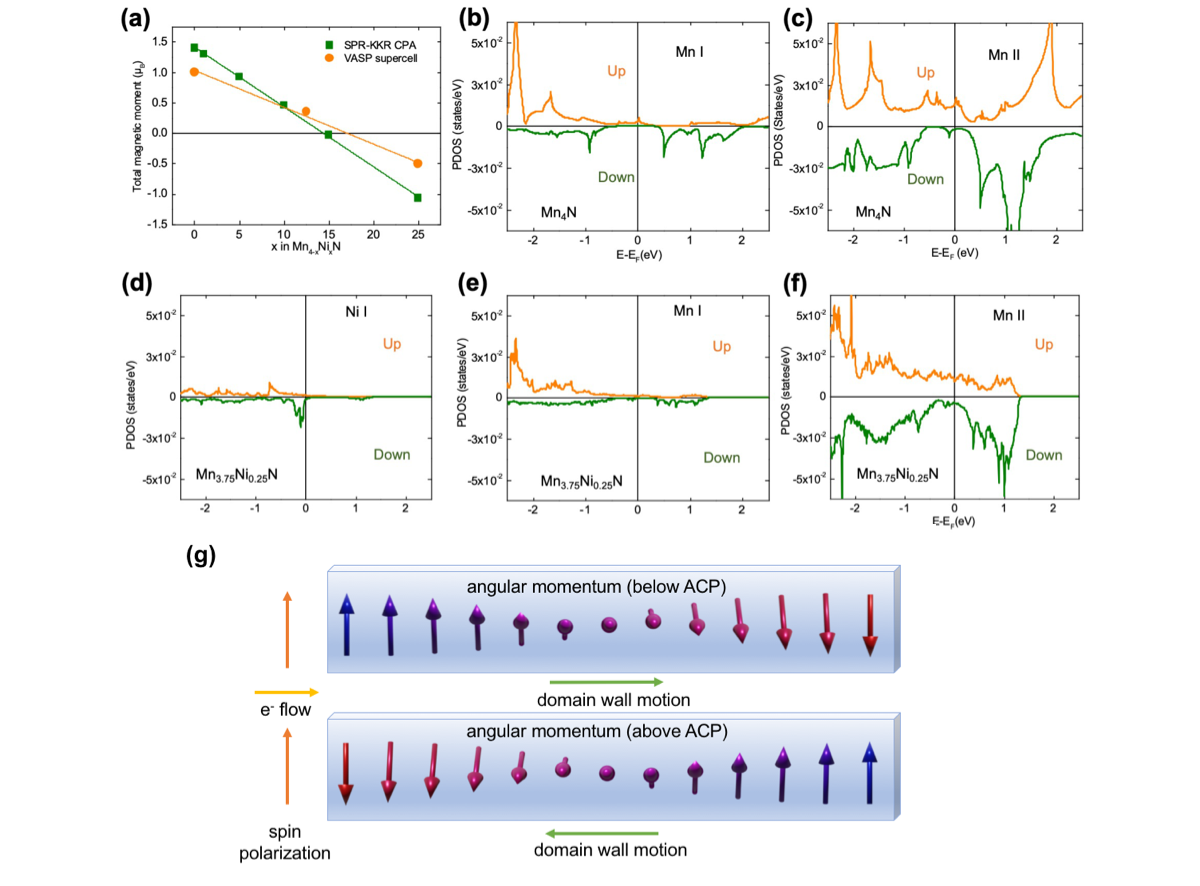}
    \caption{\textbf{(a)} Ab initio calculation of magnetic moment versus Ni concentration showing a change in the net magnetization direction around x=0.15. \textbf{(b)} \& \textbf{(c)} s-orbital PDOS of sites Mn(I) and Mn(II) of Mn$_4$N. The polarization direction at the Fermi level is "up" for both atoms while a much larger PDOS, determining the conduction electron carriers,  is obtained for the Mn(II) site. \textbf{(d)},\textbf{(e)} \& \textbf{(f)} s-orbital PDOS of Ni, Mn(I) and Mn(II) in Mn$_{3.75}$Ni$_{0.25}$N. The polarization direction at the Fermi level remains "up"  for all the atoms. \textbf{(g)} Sketch of a Bloch domain wall with the individual net magnetic moments, for a sample below/above  the angular moment compensation point (top/bottom). The orange arrows indicates the direction of the spin polarization, that stays the same below and above the ACP. The yellow line indicates the direction of the electron flow.  Since the spin polarization direction is the same, while the net   angular momentum changes direction, the spin-transfer torque drives the DW in opposite directions below and above the ACP.}
    \label{fig:4}
\end{figure}

Fig.~\ref{fig:4}a shows that the calculated total magnetic moment decreases linearly with the Ni concentration and reverses direction with respect to the global quantization axis for $x$=0.15 using the SPR-KKR approach and for $x$=0.17 using the VASP approach, in good agreement with the experimental results. Fig.\ref{fig:4}b,c present the projected density of states  (PDOS) calculated for the Mn(I) and Mn(II) sites of Mn$_4$N, while Fig.~\ref{fig:4}d,e,f show the s-PDOS of Ni substituted at site Mn(I), together with that of the Mn(I) and Mn(II) sites of Mn$_{3.75}$Ni$_{0.25}$N. For both samples the PDOS at the Fermi level ($E_F$) is about one order of magnitude larger for the Mn(II) site, with respect to that of the Mn(I) and the Ni sites. Moreover, it is larger for the majority (i.e. parallel to the global quantization axis) than the minority electrons. In summary, for compositions both below and above the ACP and MCP, the spin polarization of the conduction electrons is parallel to the (001) quantization axis and mostly due to the majority carriers in Mn(II) atoms. These  results show that the sign of the polarization is the same below and above the ACP, confirming that our experimental results are in agreement with the predictions of the $q-\phi$ model (Equation 1). 
 
 \section*{Conclusions}

In conclusion, we have investigated Mn$_{4-x}$Ni$_x$N ferrimagnetic thin films in which magnetization and angular momentum  compensation can be obtained by a fine tuning of Ni doping.  Current-driven domain wall velocity measurements were carried out for samples with various Ni content, having net magnetization on either sides of the compensation point.
The domain wall velocities, driven by STT, are observed to increase for Ni compositions close to the ACP, that in these samples coincides with the MCP. 
Record domain wall velocities approaching 3000 m/s are observed for relatively low current densities ($J$= 1.2 x 10$^{12}$ A/m$^2$). These velocities, measured at room temperature and without the support of an in-plane magnetic field  \cite{Caretta2020},  are of the same order of magnitude of those observed recently in ferrimagnetic thin films where DWs walls are driven by SHE-SOT \cite{caretta_fast_2018,Caretta2020}.  

The domain walls move in opposite directions for compositions on either sides of the ACP. This is  related to the change of relative orientation of the net spin polarization and the angular momentum, when the ACP is crossed. Our results are in agreement with the predictions of the $q-\phi$ model, expanded to a ferrimagnetic  system using effective parameters. \emph{Ab initio} calculations validate the theoretical treatment by showing that the spin polarization of the conduction electrons is mainly due to majority carriers in the Mn(II) sub-lattice. The motion of DW parallel/antiparallel to the electron flow for composition below/above the compensation point is in agreement with this finding. 

This work shows that although most of the recent efforts have focused on SOT-driven dynamics of DWs in thin films with DMI, STT appears to be also a very efficient way to drive DWs, provided that the film has a reduced magnetization (e.g. it is a ferrimagnet close to compensation points) and a  relatively strong spin polarization.  Our material, composed of abundant elements, and free of critical elements such as Co, rare earths and heavy metals, is a promising candidate for sustainable spintronics applications.

\section{Methods}
\subsection{Sample growth} 
The magnetic films were deposited on commercial 300~$\mu$m STO(100) substrates. The substrates were treated with HF and NH$_4$F solutions to smoothen the surface and were cleaned to remove any other impurities. 10~nm and 30~nm thick Mn$_{4-x}$Ni$_x$N films were then epitaxially grown at 450$\degree$ C with Mn and Ni atoms coming from the solid sources of high temperature Knudsen cells and using a radio-frequency nitrogen plasma source. The growth conditions were optimized at 1 nm/min with a N$_2$ gas flow of 0.9 cm$^3$/min with 4.1 $\times$ 10$^{-3}$ Pa  pressure in the chamber. 
To prevent further oxidation, the thin films were then capped in-situ with 3 nm of SiO$_2$ using a sputtering gun and an argon plasma source. 

\subsection{Ab initio calculations}
The Vienna ab initio simulation package (VASP) \cite{kresse_ab_1993,kresse_efficiency_1996,kresse_efficient_1996} was used for structure optimization, where the electron-core interactions are described by the projector augmented wave method for the potentials \cite{blochl_projector_1994}, and the exchange correlation energy is calculated within the generalized gradient approximation (GGA) of the Perdew-Burke Ernzerhof form \cite{perdew_generalized_1996,kresse_ultrasoft_1999}. The cutoff energies for the plane wave basis set used to expand the Kohn-Sham orbitals were 500~eV for all calculations. Structural relaxations and total energy calculations were performed ensuring that the Hellmann-Feynman forces acting on ions were less than 10$^{-2}$ N.  Using a  44x44x44~\AA$^{-1}$ k-mesh the obtained bulk lattice constant of Mn$_4$N after full lattice relaxation was 3.74~\AA. 
 Ni doping in Mn$_{4-x}$Ni$_{x}$N were taken into account using the supercell approach, where we replaced an atom of Mn I by Ni in $1 \times 1 \times 4$ and $1 \times 1 \times 8$ unit cell to model $x$ = 0.25 and $x$ = 0.125, respectively. To verify the supercell approach we also calculated the effect of Ni doping in Mn$_{4-x}$Ni$_x$N using the coherent potential approximation as implemented in SPR-KKR code.


\subsection{Domain wall velocity measurements}
Polar magneto-optical Kerr microscopy was used to image the magnetic structure. Strong out-of-plane magnetic field pulses were used to nucleate reverse domains in the nucleation pads and to inject the domain walls into the nanowires. The domain walls were displaced  by injecting both positive and negative 1~ns long current pulses. The shape of the pulses was captured and stored using an oscilloscope. 
The DW displacements from the different wires were averaged to obtain a precise estimation of the domain wall velocity. This was obtained by dividing averaged displacement by the averaged full width at half maximum (FWHM) of the pulse widths multiplied by the number of pulses. 



\section{Acknowledgements}
We acknowledge Dr. Isogami from NIMS in Japan for the measurement of the damping factor. The devices were prepared in PTA platform from Grenoble, with partial support from the French RENATECH network. We acknowledge funding from IDEX-DOMINO project and JSPS KAKENHI (No. 19KK0104 and 19K21954). J.P.G. acknowledges the European Union’s Horizon 2020 research and innovation program under Marie Sklodowska-Curie Grant Agreement No. 754303 and the support from the Laboratoire d'excellence LANEF in Grenoble (ANR-10-LABX-0051).

\section{Author Information}

\subsection{Corresponding Authors}
\renewcommand{\labelitemi}{$\textasteriskcentered$}
\begin{itemize}
    \item E-mail: stefania.pizzini@neel.cnrs.fr
    \item E-mail: laurent.vila@cea.fr
\end{itemize}

\subsection{Author Contributions}
L.V., J.P.A., T.S. and S.P. managed the project and conceived the experiments.
S.G., T.G., T.O., T.H., and H.M.  prepared the samples.
S.G. and T.G. realized the transport measurements.
S.G. and T.G. performed the domain wall dynamics measurements with the guidance of S.P.
H.O. made the TEM measurements.
J.V. and S.G. performed the magnetization measurements.
L.V., T. G. and S.G. patterned the samples by electron beam lithography.
J.P.G. applied the analytical model to the experimental results. 
A. H. and M.C. made the ab initio calculations.
S.P., J.P. A. and S.G. wrote the manuscript.
All the authors discussed the results and commented on the manuscript.



\pagebreak
\newpage

\begin{figure}[htbp]
    \centering
    \includegraphics[scale=0.8]{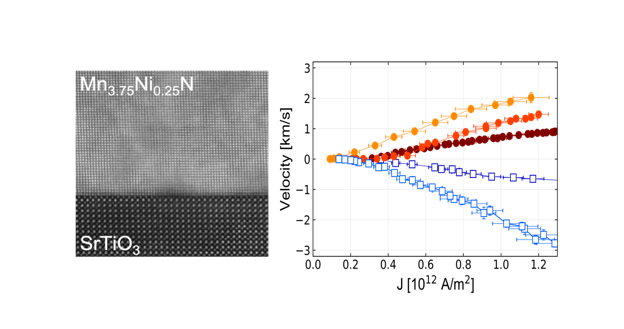}
     \caption{Figure TOC}
    \label{fig:TOC}
\end{figure}
\end{document}